\documentclass[10pt,tightenlines,
nofootinbib,prd,
showpacs,preprintnumbers]{revtex4}

\usepackage{epsfig}

\newcommand{\beq}{\begin{equation}} 
\newcommand{\eeq}{\end{equation}} 
\newcommand{\bea}{\begin{eqnarray}} 
\newcommand{\eea}{\end{eqnarray}} 
 
\newcommand{\junk}[1]{} 
\def\<{\langle} 
\def\>{\rangle} 
\def\d{\partial} 
\def\+{\dagger} 
\def\UA{$U(1)_{A}~$}
\def\UB{$U(1)_{B}~$}
\def\UY{$U(1)_{Y}~$}
\def\mueff{\mu_{\mathrm{eff}}}
\def\mkzero{m_{K^0}}

\def\kzero{K^0}
\def\kplus{K^+}
\def\vpi{v_{\pi}}
\def\fpi{f_{\pi}}
\def\thetak{\theta_{K^0}}

\def\thetacritone{53^o}

\def\thetacrit{\theta_{\mathrm{crit}}}
\def\rtil{\tilde{r}}
\def\trace{\mathrm{Tr}}
\def\diag{\mathrm{diag}}

\begin{document} 
 
\title{ Superconducting $K$ strings in high density QCD} 
\author{Kirk~B.W.~Buckley and Ariel~R.~Zhitnitsky} 
\affiliation{Department of Physics and Astronomy, 
University of British Columbia,
Vancouver, BC V6T 1Z1, Canada  }
\begin{abstract} 
Recently it has been argued that the ground state of high 
density QCD is likely be a combination of the CFL-phase along 
with condensation of the $K^0$ field. This state spontaneously 
breaks a global \UY symmetry, therefore one would expect the 
formation of \UY global strings. We discuss the core structure 
of these strings and demonstrate that under some conditions the 
global \UY symmetry may not be restored inside the string, in 
contrast with the standard expectations. Instead, $\kplus$ 
condensation occurs inside the core of the string if a relevant 
parameter $ \cos\thetak \equiv {\mkzero^2}/{\mueff^2}$ is 
 larger than some critical value $\thetak \geq \thetacrit$. If 
this phenomenon happens, the \UY strings become superconducting 
and may considerably influence the magnetic properties of dense 
quark matter, in particular in neutron stars.
\end{abstract} 
 
 \maketitle 
\section{Introduction}   

It is generally believed that
there are no   topological defects within the Standard Model in vacuum.
However, it has been realized quite recently
\cite{ssz,sonKaon,fzstrings,krstrings,kirk} that some topological defects 
such as domain walls and  strings or vortices may exist within the Standard 
Model in an unusual environment such as at large baryon density. 

In the last few years there has been a renewed interest in high 
density QCD. Similar to the BCS pairing in conventional superconductivity, 
the ground state of QCD at high density is unstable due to the formation
of a diquark condensate \cite{baillov,colorsc,colorsc2} 
(see \cite{cscqcdrev} for a review). 
In this new ground state various symmetries, which were present 
at zero baryon density, are spontaneously 
broken. Specifically, we discuss in this paper
the three flavor $N_f=3$ color flavor locking (CFL) phase. In this case
both the  exact \UB and the approximate axial \UA  symmetries of QCD are 
spontaneously broken, leading
to the formation of \UA and \UB global strings described in \cite{fzstrings}. 
The  main global property of the strings, the string tension $\alpha$, 
has been calculated in \cite{fzstrings} with the result 
$\alpha\sim 2\pi v_{\pi}^2 f_{\pi}^2\ln R$, 
where $f_{\pi}\sim \mu$ is the corresponding decay constant  in 
dense QCD set by the  chemical potential $\mu$; $v_{\pi}^2\simeq 1/3$ 
is the velocity of Goldstone modes in this media, 
and $R$ is an upper cutoff determined by the environment of the string 
(for example the presence of other strings).

If, in addition to the CFL phase,  kaon condensation also occurs as argued 
in \cite{schafer,bedascha,kaplredd,bedaque} then the hypercharge
symmetry \UY is also spontaneously broken. If this is indeed the case, one 
more type of strings related to the spontaneously broken \UY global 
symmetry is possible as was discussed in \cite{fzstrings}. 
The string tension $\alpha$ in this case behaves similarly to \UA 
and \UB global strings, and it is determined by the corresponding decay 
constant $f_{\pi}\sim \mu$. The next step in studying of \UY strings 
was undertaken in \cite{krstrings} where
it was demonstrated that  the internal   core  structure of the  
\UY string could be very different from the \UA and \UB global strings 
described earlier \cite{fzstrings}. 
Namely, it was argued that the relevant symmetry may not be restored
in the core: in most known cases, in particular in magnetic 
vortices in a conventional superconductor, the $U(1)$ symmetry is restored 
inside the core. If this is the case, the \UY string 
becomes superconducting with the core having a $K^+$ condensate. 
The fact that such unusual behavior, in principle,  may occur in the
theory of superconducting cosmic strings and in quantum field
theory in general,  has been known for quite a while \cite{witten},
\cite{turner}, \cite{core}. Even more than that, such a behavior has been 
observed experimentally when the laboratory experiments on $^3He$ provided 
us with strong evidence for defect core transition in the interior of 
vortices which appear in the superfluid $^3He -B$ phase 
(see \cite{He} for a review). 
Still, such a behavior of the vortex core is considered as an exception
rather than a common phenomenon in physics.

Due to the fact that the \UY strings might be phenomenologically relevant 
objects realized in nature (presumably in neutron stars),  and due
to the property  of  superconductivity  of these strings which
might be  relevant for the dynamics, we analyze the core 
structure of \UY strings in detail in this letter. 
More specifically, the goal of this letter is twofold: 
1) understand  the phenomenon of core transition in the interior of 
vortices qualitatively, using some analytical methods; 2) make quantitative
estimates for phenomenologically relevant parameters in the CFL phase when 
the transition does  occur.

The fact that there should be some kind of phase transition 
in the string core as a function of the external parameters
can be understood from the following simple  arguments. 
If the quark mass difference $m_d-m_u$ is relatively  large, than 
there would be only a $U(1)$ (rather than $SU(2)_I$) symmetry broken. The 
standard topological arguments suggest that in this case, the  $K^0$ string 
would be a topologically stable configuration with the restoration of 
the corresponding $U(1)$ symmetry inside the core (which is a 
typical situation). If  $m_d-m_u$ is exactly zero, such that the isotopical
$SU(2)_I$ symmetry is exact, then symmetry arguments suggest that  
both $K^0$ and $K^+$ fields condense, and no global stable strings 
are possible. From these two limiting cases, it is clear that 
there should be some intermediate region that somehow interpolates 
(as a function of $m_d-m_u$)  between these two cases.
Indeed, as we discuss below in detail, the way how this interpolation works
is the following. For relatively large $m_d-m_u$ nothing unusual happens: 
the $K^0$ string has a typical behavior with $\kzero$ condensation  
outside the core, and  with restoration of the symmetry inside the core.
At some finite magnitude of    $m_d-m_u$, an instability arises through 
the condensation of $\kplus$-field inside of the core of the string. 
As the magnitude of $m_d-m_u$ decreases,  the size of the core 
becomes larger and larger
with nonzero values of both $K^0$ and $K^+$ condensates inside the core.
Finally, at $m_d=m_u$ the core of the string 
(with nonzero condensates $K^0$ and $K^+$)
fills the entire space, in which case the meaning of the string 
is completely lost, and we are left with the situation when $SU(2)_I$ 
symmetry is exact: no stable strings are possible.

Given this argument, we would expect that there must be some transition region
where the $SU(2)_I$ symmetry is broken to some degree below which
$\kplus$-condensation will occur inside the core. The point at which this 
occurs will be estimated in this letter. We will show that 
$\kplus$-condensation only occurs above a certain point 
$\thetak > \thetacrit$, where  parametrically $\thetacrit$ is given by  
$\sin(\thetacrit/2)\simeq {\mathrm{constant}}~ 
(\Delta/m_s)\sqrt{(m_d-m_u)/m_s}$, 
with $\Delta\sim 100$ MeV being the superconducting gap and $m_s$ 
is the strange quark mass.
 
This paper is organized as follows. In section II we will give a brief 
overview of the properties of the mixed CFL-$\kzero$ phase of high
density QCD.  Section III will discuss the issue of the stability
of global $\kzero$ strings
as a function of the parameter $(m_d-m_u)$.  In this section we will calculate 
$\thetacrit$ where $K^+$ condensation occurs inside of the global strings. 
We end with concluding remarks. 
 
 
\section{The CFL+$\kzero$ phase of high density QCD}  
It is well known that the ground state of $N_f=3$, $N_c=3$ QCD exhibits  
the Cooper pairing phenomenon as in conventional superconductivity 
\cite{colorsc,colorsc2,CFL,rapp}. 
The corresponding  condensates in the CFL phase take the approximate form:
\bea 
\label{qqCFL} 
  \<q^{ia}_{L\alpha} q^{jb}_{L\beta} \>^* &\sim&   
  \epsilon_{\alpha\beta\gamma} \epsilon^{ij}\epsilon^{abc} X_c^{\gamma} , 
  \nonumber \\ 
  \<q^{ia}_{R\alpha} q^{jb}_{R\beta} \>^* &\sim&   
  \epsilon_{\alpha\beta\gamma} \epsilon^{ij}\epsilon^{abc}Y_c^{\gamma} , 
\eea 
where $L$ and $R$ represent left and right handed quarks, $\alpha$, $\beta$, 
and $\gamma$ 
are the flavor indices, i and j are spinor indices, a, b, and c are  
color indices, and $X_c^{\gamma}$ and $Y_c^{\gamma}$ 
are complex color-flavor matrices describing the Goldstone bosons.
In order to describe the light degrees of freedom in a gauge invariant way,  
one introduces the color singlet field $\Sigma$
\beq 
\label{compsigma}
\Sigma_{\gamma}^{\beta}= X Y^{\+}  = \sum_c X_c^{\beta} Y^{c *}_{\gamma},
\eeq 
such that the leading terms of the effective Lagrangian in terms of $\Sigma$
take the form \cite{bedascha}
\bea
\label{Leffsigma}
{\cal L}_{\mathrm{eff}}=\frac{\fpi^2}{4} \trace \left[
        \nabla_0\Sigma \nabla_0\Sigma^{\+} 
        - \vpi^2 \d_i \Sigma \d_i \Sigma^{\+} \right] 
        + 2 A \left[ \det(M) \trace(M^{-1} \Sigma + h.c. \right] ,\\
\nabla_0 \Sigma = \d_0 \Sigma 
        + i \left(\frac{M M^{\+}}{2 p_f} \right) \Sigma
        - i \Sigma \left(\frac{M^{\+} M}{2 p_f}  \right) \nonumber,
\eea
where the matrix $\Sigma= \exp(i \pi^a \lambda^a/ \fpi )$ describes the 
octet of 
Goldstone bosons with the $SU(3)$ generators $\lambda^a$ normalized as 
$\trace(\lambda^a \lambda^b)=2 \delta^{ab}$. 
The quark mass matrix in Eq.(\ref{Leffsigma})
is given  by $M=\diag(m_u,m_d,m_s)$. Finally, we neglect the electromagnetic 
interactions in the expression (\ref{Leffsigma}) but keep the isospin 
violation $\sim (m_d-m_u)$, which is an appropriate approximation 
for the physically relevant case when the  baryon density is not 
very high \cite{bedaque}. The constants $\fpi, \vpi$ and
$A$ have been calculated in the leading perturbative approximation and 
are given by \cite{ss,bbs}:
\beq
\fpi^2=\frac{21-8 \ln 2}{18} \frac{\mu^2}{2 \pi^2}, ~~~
\, \vpi^2=\frac{1}{3},~~~  \, A=\frac{3 \Delta^2}{4 \pi^2}.~~~ 
\eeq

Recently, it was realized in 
\cite{bedascha,bedaque,schafer,kaplredd} that the ground state
of the theory may be different from the pure CFL phase for 
a physical value of the strange quark $(m_s \gg m_u,m_d)$: condensation of 
the $\kzero$ and $\kplus$ mesons would lower  the free energy of the system
by reducing the strange quark content. Specifically, it has been  argued that 
kaon condensation would occur in the CFL phase if $m_s \geq 60$ MeV.  
This means that $\Sigma ={\bf 1}$ is no longer the global minimum of the 
free energy; instead, some rotated value of $\Sigma$ describes the ground 
state in this case. In what follows we consider the realistic case 
when the isospin symmetry is not exact, $m_d> m_u$, such that 
$\kzero$ condensation occurs. The appropriate
expression for $\Sigma$ describing the $K^0$ condensed ground state 
in this case can be parameterized as:
\beq
\Sigma= \left( \begin{array}{ccc}
     1 & 0 & 0 \\
     0 & \cos \thetak & \sin \thetak e^{-i\phi} \\
     0 & -\sin \thetak e^{i\phi} & \cos \thetak
 \end{array} \right), 
\eeq
where $\phi$ describes the corresponding Goldstone mode 
and $\thetak$ describes the strength of the kaon condensation with 
\cite{bedascha}:
\beq
\label{vevtheta}
\cos \thetak = \frac{m_0^2}{\mueff^2}, ~~~ \mueff\equiv\frac{m_s^2}{2p_F}
,~~~~m_0^2\equiv a m_u (m_d + m_s), ~~~a=\frac{4A}{f_{\pi}^2}
=\frac{3 \Delta^2}{\pi^2 \fpi^2}.~~~  
\eeq
In order for this to be satisfied, we must have $ m_0 < \mueff$. If kaon 
condensation occurs and $\thetak\neq 0$,  an additional $U(1)$ symmetry is 
broken. This brings us to the next section where 
will discuss the consequences of this symmetry breaking. 


\section{Global $K-$ strings}
We follow our logic outlined in the Introduction and first consider
global $\kzero$ strings when they are topologically stable. 
This corresponds to the  approximation  
when the splitting between $K^0$ and all other degrees of freedom is relatively
large such that  we can neglect   in our effective description
all fields except $\kzero$.  After that we analyze situation when the 
$K^0$ and $K^+$
masses are degenerate such that the $K$ strings become unstable. Finally we 
introduce a small explicit isospin violation into our description
  $\sim (m_d-m_u)$ in order to analyze the stability/instability issue for this
physically relevant case.
The  important global characteristic of the $K^0$ string, the string 
tension $\alpha$,
with logarithmic accuracy is determined by the pion decay constant 
$\alpha \sim f_{\pi}^2$ as discussed in \cite{fzstrings},
and it is not sensitive to the internal structure of the core. The subject
of this paper is the analysis of the  core structure of the $\kzero$ strings.

\subsection{Topologically stable $U(1)$ strings in $CFL+K^0$ phase}

 We start by considering the
following effective field theory which describes a single complex $K^0$ 
field. This corresponds to the 
case of spontaneously broken $U(1)$ symmetry 
\beq
\label{K1}
{\cal L}_{\mathrm{eff}}(\kzero)
  =|(\d_0+i \mueff)\kzero|^2 - \vpi^2 |\d_i \kzero|^2
  - m_0^2 |\kzero|^2 - \lambda |\kzero|^4 .
\eeq
We   fix all parameters of this effective theory by comparing the amplitude
of the $\kzero$ field to the result obtained from the    theory described in 
the previous section. We neglect all other degrees of freedom at this point 
(see the discussion at the end of this section).
If $\mueff > m_0$ the kaon field acquires a nonzero vacuum expectation value 
$\<\kzero\>=\eta/\sqrt{2}$ where 
\beq
\label{K2}
\eta^2=\frac{\mueff^2 -m_0^2}{\lambda} 
     =\frac{\mueff^2}{\lambda} \,(1 - \cos \thetak).
\eeq
For $\mueff > m_{\kzero}$ it is more convenient
to represent the effective Lagrangian in the familiar
form of a Mexican hat type potential:
\beq
\label{K3}
{\cal L}_{\mathrm{eff}}=|\d_0 \kzero|^2 - \vpi^2 |\d_i \kzero|^2
  - \lambda \left(|\kzero|^2 - \frac{\eta^2}{2}\right)^2 .
\eeq
This is a text-book Lagrangian with spontaneously broken global $U(1)$ symmetry
which admits topologically stable global string solutions. 
As is well-known,  
global strings are solutions of the time independent equation of motion. 
The time independent equation of motion for $\kzero$  
is given by:
\bea
\label{K4}
\vpi^2 \nabla^2 \kzero &=& 2 \lambda \left(  
 |\kzero|^2 - \frac{\eta^2}{2}\right) \kzero .
\eea
For the $\kzero$ string solution we will make the following ansatz:
\bea
\label{k0string}
\kzero_{\mathrm{string}}&=&\frac{\eta}{\sqrt{2}} f(r) e^{in \phi},  
\eea 
where $n$ is the winding number of the string (we will take $n=1$ in what 
follows), $\phi$ is the azimuthal angle in cylindrical coordinates, and 
$f(r)$ is a yet to be determined solution of Eq. (\ref{K4}) which obeys 
the boundary conditions $f(0)=0$ and $f(\infty)=1$. Substituting this 
ansatz into Eq. (\ref{K4}), we arrive at the following ordinary 
differential equation:
\beq
\label{K5}
\frac{1}{r} \frac{d}{dr} \left(r \frac{df(r)}{dr} \right) - \frac{f(r)}{r^2}
=\frac{\mueff^2(1 - \cos \thetak)}{\vpi^2} (f(r)^2-1) f(r),
\eeq
where we replace $\lambda\eta^2\rightarrow \mueff^2(1 - \cos \thetak)$
according to Eq. (\ref{K2}).
Although a numerical solution of this equation is possible,
the most important part of this section is an analytical analysis and, 
therefore, we prefer to  take a 
variational approach. We  follow \cite{turner,brandpistring} 
and assume a solution of the form:
\beq
f(r) = 1 - e^{-\beta r} ,
\eeq
with the variational parameter $\beta$. Minimizing the energy with 
respect to $\beta$, the result is \cite{brandpistring}:
\beq
\label{betamin}
\beta^2_{\mathrm{min}}
=\frac{89}{144} \frac{\mueff^2}{\vpi^2}(1- \cos \thetak).
\eeq
This result was compared with the exact numerical solution of (\ref{K5})
and is a reasonable approximation. From this equation  we see that a typical  
string core  radius is parametrically
 given by  $r_{\mathrm{core}} \sim 1/\beta_{\mathrm{min}} \sim 
\vpi/ \mueff( \sin \thetak/2)$,and it becomes smaller
when $\mueff$   gets larger. The first important lesson of this simple exercise
is the observation that the global string is stable and the $U(1)$ symmetry is 
restored inside of the core, $f(r=0)=0$, as expected.    
The second important lesson is the observation that the core size 
becomes large when $\thetak\rightarrow 0$.

\subsection{Unstable $ SU(2)_I \times U(1)_Y\rightarrow U(1)$ strings in 
$CFL/K^0$ phase} 

Our next task is an analysis of the $K$ string when 
isospin is an exact symmetry (  i.e. $m_u=m_d$), such that 
the symmetry pattern breaking is
$ SU(2)_I \times U(1)_Y\rightarrow U(1)$. In this case, based on topological
arguments, we know that the $K$ string is unstable. We want to analyze the 
stability issue in detail to understand this phenomenon on the 
quantitative level. 

To simplify things we start with the effective Lagrangian
which only includes a single  complex kaon doublet $\Phi=(\kplus,\kzero)$.
As discussed in \cite{bedascha}, this is an appropriate
approach  to discuss kaon condensation in the CFL phase
if  all other degrees of freedom  are much heavier. As before (see Eqs.
(\ref{K1},\ref{K3})) we can represent the effective Lagrangian 
in the form of a Mexican hat type potential,
\beq
\label{Leff}
{\cal L}_{\mathrm{eff}}(\Phi)
=|\d_o \Phi|^2 - \vpi^2 |\d_i \Phi|^2
  - \lambda \left(|\Phi|^2 - \frac{\eta^2}{2}\right)^2 .
\eeq
All parameters here  are defined in the same way as in 
(\ref{vevtheta},\ref{K2},\ref{K3}) and the
$\Phi$ field acquires a non-zero vacuum expectation value
which we assume takes the form
\beq
\label{kzerovev}
\< \Phi \> = (0, \frac{\eta}{\sqrt{2}}).
\eeq 
The time independent equations of motion for $\kzero$ and $\kplus$ 
are given by:
\bea
\label{eomkplus}
\vpi^2 \nabla^2 \kplus &=& 2 \lambda (|\kplus|^2 
 + |\kzero|^2 - \frac{\eta^2}{2}) \kplus ,\\
\label{eomkzero}
\vpi^2 \nabla^2 \kzero &=& 2 \lambda (|\kplus|^2 
 + |\kzero|^2 - \frac{\eta^2}{2}) \kzero .
\eea
For the $\kzero$ string solution we will make the following ansatz:
\bea
\label{k0string2}
\kzero_{\mathrm{string}}&=&\frac{\eta}{\sqrt{2}} f(r) e^{in \phi},  ~~~
\kplus = 0, 
\eea 
such that $f(r)$ is the solution of  Eq. (\ref{K5}) with
the boundary conditions $f(0)=0$ and $f(\infty)=1$.  Without calculations,
from topological arguments we know that although the solution
(\ref{k0string2}) satisfies the equation of motion (\ref{eomkplus},
\ref{eomkzero}), it is an unstable solution.
The source of instability can be seen as follows.
We follow the standard procedure and  expand the energy in the   
$\kzero$ string background to quadratic order in $\kplus$ and $\kzero$ modes:
\beq
E  (\kzero=\kzero_{\mathrm{string}}+\delta\kzero, ~~\kplus)
\approx E_{\mathrm{string}} + \delta E.
\eeq
We know that the $\kzero$ string itself is a stable configuration, therefore, 
the $\delta\kzero$ modes cannot have negative eigenvalues which 
would correspond to the instability.
Therefore, we concentrate only on the ``dangerous'' modes related to $\kplus$ 
fluctuations, in which case $\delta E$ is given by:
\beq
\label{K6}
\delta E = \int d^2r \left( \vpi^2 | \nabla \kplus |^2 
        + \mueff^2(1-\cos\thetak)(f(r)^2-1) |\kplus|^2 
         \right), 
\eeq
where $f(r)$ is the solution of  Eq. (\ref{K5}) with
the boundary conditions $f(0)=0$ and $f(\infty)=1$.
If $\delta E$ is a positive quantity, then the $\kzero$ string is absolutely
stable and $\kplus$  modes do not destroy the string configuration. 
If $\delta E$ is negative, this means that this is a direction in 
configurational space where the $\kzero$ string decays.
Following \cite{vachaspati}, the $\kplus$ field can be expanded in 
Fourier modes:
\beq
\label{fourierexp}
\kplus = \frac{\eta}{\sqrt{2}} \sum_m g_m(r) e^{i m \phi}. 
\eeq
Now we have the $\kplus$ field in terms of the dimensionless Fourier
components $g_m(r)$. 
Setting $m=0$ in the above expansion in order to analyze the lowest energy 
$\delta E_0$ contribution in (\ref{K6}), we arrive at:
\beq
\label{K7}
\delta E_0 = \frac{\eta^2}{2}\int d^2r 
	\left( \vpi^2 (\frac{\d g_0(r)}{\d r})^2 
        + \mueff^2(1-\cos\thetak)\cdot (f^2(r)-1) g_0^2(r) 
         \right) .
\eeq
In order to have dimensionless coordinates and fields, we will perform the
following change of variables, $\rtil=\gamma r$, 
where $\gamma \equiv \mueff/ \vpi$.
This change of variables sets the string width
in $f(\rtil)$ to be $\rtil_{\mathrm{core}} \sim 1$. 
Equation (\ref{K7}) now reads:
\bea
\label{K8}
\delta E = \frac{\eta^2 \vpi^2}{2}\, 
        \int d^2\rtil\,g_0(\rtil)\hat{O} g_0(\rtil) ,~~~
        \hat{O} = - \frac{1}{\rtil} \frac{d}{d \rtil}(\rtil \frac{d}{d \rtil}) 
        + \lambda' (f^2(\rtil) -1) , ~~~\lambda'= (1 - \cos \thetak).
\eea
The problem is reduced to the analysis of the 
two-dimensional Schrodinger equation for a particle in an attractive potential 
$V(r)=-(1 - \cos \thetak)(1-f^2(\rtil))$ with $f(\rtil)$ being the solution
of  Eq. (\ref{K5}) with
the boundary conditions $f(0)=0$ and $f(\infty)=1$.
Such a potential is negative everywhere and approaches zero at infinity.
As is known from standard quantum mechanics \cite{Landau},
for an arbitrarily  weak potential well there is always a negative 
energy bound state in one and two spatial dimensions; in three
dimensions a negative energy bound state may not exist. For the 
two dimensional case (the relevant problem in our case) the lowest energy 
level of the bound state is always negative and exponentially small for small 
$\lambda'$. One should note that our specific potential 
$V(r)=-(1 - \cos \thetak)(1-f^2(\rtil))$ which enters (\ref{K8})
is not literally the potential well, however one can always construct the 
potential well $ V' $ such that its absolute value is smaller than  
$ | V(r) | $ everywhere, i.e. $ | V' | < | V(r) | $
for all $ r $. For the potential well $V'$ we know that 
the negative energy bound state always exists; 
when $V'$ is replaced by $V$ it makes the energy eigenvalue even lower.
Therefore, the operator (\ref{K8}) has always a negative mode irrespective of 
the local properties of function $f(r)$.
As a consequence, the string (\ref{k0string}) is an unstable solution of the 
classical equation of motion, the result we expected from the
beginning from topological arguments. The instability manifests itself 
in the form of a negative energy bound state solution of the corresponding 
two-dimensional Schrodinger equation (\ref{K8}) irrespective of the 
magnitudes of the parameters.

\subsection{$\kplus$-condensation in the core of $\kzero$ strings in 
$CFL+K^0$ phase} 

The issue of the stability or instability  of $\kzero$ strings reviewed in the 
previous section is highly sensitive to the degree of symmetry present in 
the Lagrangian describing the $K^0/K^+$ system. If the $SU(2)_I$ symmetry is 
strongly  broken, the $\kzero$ strings will be absolutely 
(topologically) stable 
as discussed in subsection  A. If the $SU(2)_I$ symmetry remains unbroken,   
the $\kzero$ strings will always be unstable 
as discussed in subsection B. Now we introduce an explicit symmetry breaking 
parameter $\delta m^2$ into (\ref{Leff}) fixed by the original Lagrangian 
(\ref{Leffsigma}) such that our simplified version of the system 
(only $\Phi =(K^+, K^0)$ fields are taken into account) has the form
\footnote{In addition to the mass splitting proportional to the difference
$m_d-m_u$, there is also a splitting due to electromagnetic effects, 
$\delta m_{EM}^2 \sim \alpha_{EM} \Delta^2/ (8 \pi)$. 
As mentioned in section 2, the electromagnetic contribution becomes
important at very large 
chemical potential \cite{bedaque}. However, this correction can be neglected 
for the present work since we are not considering large chemical potentials
and it only amounts to a 10\% 
correction, $\delta m_{EM}/\delta m \sim 0.1$. In order to remain 
self-consistent, we neglect all electromagnetic contributions throughout
this paper.}
\beq
\label{K9}
{\cal L}_{\mathrm{eff}}(\Phi)
=|\d_0 \Phi|^2 - \vpi^2 |\d_i \Phi|^2
  - \lambda \left(|\Phi|^2 - \frac{\eta^2}{2}\right)^2 -
\delta m^2 \Phi^{\+} \tau_3 \Phi ,~~~ \delta m^2\equiv \frac{a}{2}m_s(m_d-m_u).
\eeq
We anticipate that, as the symmetry breaking 
parameter $\delta m^2$ in Eq.~(\ref{K9}) becomes sufficiently large,
a stable $K^0$ string  with restored  $U(1)$ symmetry in the core, $f(r=0)=0$,
must be reproduced, as discussed in subsection A. When the symmetry breaking 
parameter $\delta m^2$ in becomes sufficiently small,
one should  eventually reach a point where $K^+$ instability
occurs, and it is energetically favorable for a $\kplus$
condensate to be formed inside the core of the string.

In this subsection we  calculate the critical value of $\thetak$ when the 
$\kplus$ instability occurs, and a $\kplus$ condensate does 
form in the string core. In addition to this, we will obtain an estimate 
of the absolute value of the $\kplus$-condensate at the center of the core 
of the string $(r=0)$. In order to determine if $\kplus$-condensation 
occurs within the core of $\kzero$-strings, we would like 
(ideally) to solve the set of coupled differential equations
\bea
\label{K10}
\vpi^2 \nabla^2 \kplus &=& 2 \lambda (|\kplus|^2 
 + |\kzero|^2 - \frac{\eta^2}{2}) \kplus +\delta m^2\kplus,\\
\label{K11}
\vpi^2 \nabla^2 \kzero &=& 2 \lambda (|\kplus|^2 
 + |\kzero|^2 - \frac{\eta^2}{2}) \kzero -\delta m^2\kzero.
\eea
with the appropriate boundary conditions. 
This is not a trivial task, and we will follow a different approach.  
The main point is:  we are mainly interested in the {\it critical values}
of the parameters when $\kplus$-condensation starts to occur inside the core. 
In this case we can treat $\kplus$ field as
a small perturbation in the  $\kzero$ background field. Such an approach
is not appropriate when  $\kplus$-condensation is already well-developed 
in which case both fields $\kzero$ and $\kplus$ must be treated on the 
same footing. However, 
this approach is quite appropriate when one studies the transition  from the
phase where $\kplus$ background field is zero to the region where it becomes 
nonzero.

To begin, we will expand the energy in the constant $\kzero$ string
background to quadratic order in $\kplus$:
\beq
E \approx E_{\mathrm{string}} + \delta E,
\eeq
where $\delta E$ is given by:
\beq
\label{K12}
\delta E = \int d^2r \left( \vpi^2 | \nabla \kplus |^2 
        + \mueff^2(1-\cos\thetak)\cdot(f^2(r)-1) |\kplus|^2 
        + \frac{a}{2}m_s(m_d-m_u) \, |\kplus|^2  \right). 
\eeq
If $\delta E$ is a positive quantity, then the $\kzero$ string is 
stable and $\kplus$ condensation does not occur inside the core of the 
string. If $\delta E$ is negative, this means that it is energetically
favorable for $\kplus$ condensation to occur inside the core of the 
string.  We follow the same procedure as before keeping the most 
``dangerous'' mode to arrive at:
\beq
\label{K13}
\delta E = \frac{\eta^2}{2} \int d^2r \left( \vpi^2 (\frac{\d g_0(r)}{\d r})^2 
        + \mueff^2(1-\cos\thetak)\cdot(f^2(r)-1) g_0^2(r) 
        + \frac{a}{2}m_s(m_d-m_u)  g_0^2(r)  \right) .
\eeq
In dimensionless variables this expression can be represented as follows
\bea
\label{K14}
\delta E = \frac{\eta^2 \vpi^2}{2} \, 
        \int d^2\rtil\,g_0(\rtil)[\hat{O}+\epsilon] g_0(\rtil) ,~~~
        \hat{O} = - \frac{1}{\rtil} \frac{d}{d \rtil}(\rtil \frac{d}{d \rtil}) 
        + (1-\cos\thetak) (f^2(\rtil) -1) , ~~~\epsilon\equiv  \frac{a}{2} 
        \frac{m_s (m_d-m_u)}{\mueff^2}.
 \eea
The only difference between this expression  and Eq. (\ref{K8}) describing 
the instability of the string in case of exact isospin symmetry, 
is the presence of the term $\sim \epsilon $ in Eq. (\ref{K14}).
The problem of determining when $\kplus$-condensation occurs is now reduced to 
solving the Schrodinger type equation $\hat{O} g_0 = \hat{E} g_0$. 
From the previous discussions we know that $\hat{E}$ for the ground state 
is always negative. However, to insure the  instability  with respect 
to $\kplus$-condensation one should require a relatively large negative 
value i.e.
$\hat{E}+\epsilon < 0 $. It can not happen for arbitrary weak coupling constant
$\sim (1-\cos\thetak)$ when $\thetak $ is small. However, it does happen 
for relatively large $\thetak$. To calculate the minimal critical
value   $\thetacrit$ when    $\kplus$-condensation develops, one 
should calculate the  eigenvalue $\hat{E}$ as a function of parameter 
$\thetak$ and solve the equation $\hat{E}(\thetacrit)+\epsilon = 0 $. 
As we mentioned earlier, for very small coupling constant 
$\lambda'=(1-\cos\thetak)\rightarrow 0$ the bound state energy is 
negative and exponentially small, $\hat{E}\sim -e^{-\frac{1}{\lambda'}}$.
However, for realistic parameters of $\mu ,~\Delta,~ m_s,~ m_u,~ m_d$ the 
parameter $\epsilon$ is not very small and we expect that
in the region relevant for us the bound state energy $\hat{E}$ is 
the same order of magnitude
as the potential energy $\sim \lambda'$. In this case we estimate  $\thetacrit$
from the  following conditions $-\hat{E}(\thetacrit)\sim\lambda'\sim
(1-\cos\thetacrit)\sim \epsilon$ with the result
which can be parametrically represented as
\beq
\label{K15}
\sin\frac{\thetacrit}{2}\sim \mathrm{const}
 \frac{\Delta}{\pi f_{\pi}}\sqrt{\frac{m_s(m_d-m_u)}{\mueff^2}}
\sim \mathrm{const} \frac{\Delta}{m_s}\sqrt{\frac{(m_d-m_u)}{m_s}},
\eeq
where we have neglected all numerical factors in order to explicitly
demonstrate the dependence of $\thetacrit$ on the external parameters. 
The limit of exact isospin symmetry, which corresponds to
$m_d\rightarrow m_u $ when the string becomes unstable,  
can be easily understood
from the expression (\ref{K15}). Indeed, in the case
that the critical parameter $\thetacrit \rightarrow 0$ becomes 
an arbitrarily small number the $K^+$ instability
would develop for arbitrarily small $\thetak >0 $.
The region occupied by the $K^+$ condensate at this point is determined by the
behavior of lowest energy mode $g_0$ at large distances, 
$g_0(\rtil\rightarrow\infty)\sim\exp(-\hat{E}\rtil)$ such that a typical 
$\rtil\sim (m_d-m_u)^{-1}\rightarrow \infty$ as expected. 

In order  to make a  quantitative, rather than qualitative estimation of 
the critical value $\thetacrit$ when  $\hat{E}(\thetacrit)+ \epsilon = 0$, 
we  discretize the operator $\hat{O}$ and solve the 
problem numerically, with the boundary conditions $g_0(\infty)=0$ and 
$g_0(0)=\mathrm{constant}$. Varying the condensation angle $\thetak$, we 
see that a negative eigenvalue $\hat{E}+\epsilon < 0$ appears when 
$\thetak >\thetacrit \approx \thetacritone$. For our parameters we use
$m_u=5$ MeV, $m_d=8$ MeV, $m_s=150$ MeV, $\mu=500$ MeV, and
$\Delta=100$ MeV which gives  
$\epsilon\equiv  a m_s (m_d-m_u)/(2\mueff^2) \simeq 0.1$. 
In Fig. 1 we show a plot of $g_0(\rtil=0)$ as a function of $\thetak$. 
One can see that the transition from no $\kplus$-condensation is reached 
at about $\thetacrit \approx \thetacritone$. In Fig. 2 we plot the functions
$f(\rtil)$ (related to the $\kzero$ string by (\ref{k0string})) 
and $g_0(\rtil)$ (related to the $\kplus$ condensate by (\ref{fourierexp}))
as a function of the rescaled coordinate $\rtil$. One can see that the 
$\kplus$ condensate falls off over the same distance as the $\kzero$ string
reaches its vacuum expectation value. We should note that the solution to the 
above Schrodinger equation does not give us the overall normalization of 
the function $g_0(\rtil)$. We have estimated the overall normalization of 
$g_0(\rtil)$ by minimizing the total energy of the system using 
a variational approach. 

To conclude the section, we want to make the following comment.
We have demonstrated above
that $K^+$ condensation might occur in the core of $K^0$ strings if some
conditions are met. We also explained  how this phenomenon depends on the 
external parameters. We also  derived  the equation
$\hat{E}(\thetacrit)+\epsilon=0$,
the solution of which allows us to calculate the critical parameters 
when $K^+$ condensation starts to occur. 
All these discussions were quite general because they were
based on the symmetry and topological properties of the system rather than 
on  a specific form of the interaction.  
However, the numerical estimates presented above was derived by using   
a concrete form of   
the effective Lagrangian (\ref{K9}) describing the lightest $K^+,~K^0$ 
degrees of freedom. The question arises how sensitive our
numerical results are when the form of the potential changes.
To formulate the question in a more specific way, let us remind the reader
that, in general, the effective Lagrangian
describing the Goldstone modes can be represented in many different forms   
as long as 
symmetry properties are satisfied. The results for the amplitudes  
describing the interaction of the Goldstone particles do 
not depend on a specific representation used. 
A well-known example of this fact is the possibility
of describing  the $\pi$ meson properties by using
a linear $\sigma$ model as well as a non-linear $\sigma$ model 
(and many other models which satisfy the relevant symmetry breaking pattern ). 
The results  remain the same if one discusses the local properties
of the theory (such as $\pi\pi$ scattering  length)
when the $\pi$ meson is considered as a small  quantum fluctuation rather than
a large  background field .  It may not be the case 
when $\pi$ represents a large background field in which case
some numerical difference  between different representations of the effective
Lagrangian may occur. Roughly speaking, the source of the difference is 
an inequality
$\pi(x)\neq\sin\pi(x) $ for large global background fields such as a 
string solution which is the subject of this letter. 

Having this in mind, we repeated
similar numerical estimates discussed above for the original effective 
Lagrangian (\ref{Leffsigma}) where  Goldstone fields represented
in the exponential form rather than in form determined by the effective 
Lagrangian (\ref{K9}). As before, in these estimates we  considered  
exclusively $K^+$ modes which are the energetically lowest modes and which 
can potentially destabilize the system presented by the background field 
of the $K^0$ string. This approximation
is justified as long as the $K^+$ field is the lowest massive  excitation 
in the system when a $K^0$ condensate develops. Also, the typical size of 
the core must be larger than the inverse gap $\Delta^{-1}$, i.e.  
$r_{\mathrm{core}} \simeq \vpi/ \mueff \gg\Delta^{-1}$ in order 
to maintain color superconductivity inside the core.   
We assume this is the case.  Our numerical results suggest, that the critical 
parameters are not very sensitive
to the specifics of the Lagrangian such that $\thetacrit$ is close to 
our previous numerical estimates. Therefore,  most likely, the real world 
(with our parameters  $\thetak \simeq 70^o$ ) case corresponds 
to $\thetak > \thetacrit$ and thus, a $K^+$ condensate does develop inside the 
core of the $K^0$ strings.
 

\section*{Conclusion} 
We have demonstrated that, within the CFL+$\kzero$ phase of QCD, 
$\kplus$ condensation does occur within the core of global $\kzero$ 
strings if some conditions are met: $\thetak > \thetacrit$.
We presented two estimates for $\thetacrit$: an analytical one 
which gives a qualitative explanation of the phenomenon,
as well as numerical one for the physically relevant 
parameters realized in nature. Our results suggest that if 
a CFL+$\kzero$ phase is realized in nature, it is likely that $K^0$ 
strings form together with $\kplus$ condensation inside the core, 
in which case the strings become superconducting strings (see \cite{witten}
for a more complete description of superconducting strings and 
their properties). 
 
 It is known that the CFL+$\kzero$ phase may be realized in nature in 
neutron stars interiors and in the violent events associated with 
collapse of massive stars or collisions of neutron stars, so $K^0$ 
superconducting strings with $\kplus$ condensation inside the core
could be very important for such astrophysical phenomena. 
It has been recently argued \cite{arz}
that such conditions may also occur in early universe during the 
QCD phase transition. In this case  it might be important for cosmological 
problems, such as the dark matter problem, as well.

\section*{Acknowledgments }
We would like to thank Michael Forbes for help with some of the 
preliminary numerical work and for useful comments on this paper. We would 
also like to thank Sanjay Reddy for useful communications.
This work was supported in part by the Natural Sciences and Engineering 
Research Council of Canada.

\begin{figure}
\begin{center}
\vspace{1cm}
\epsfxsize=6.0in
\epsfbox[ 61 199 549 589]{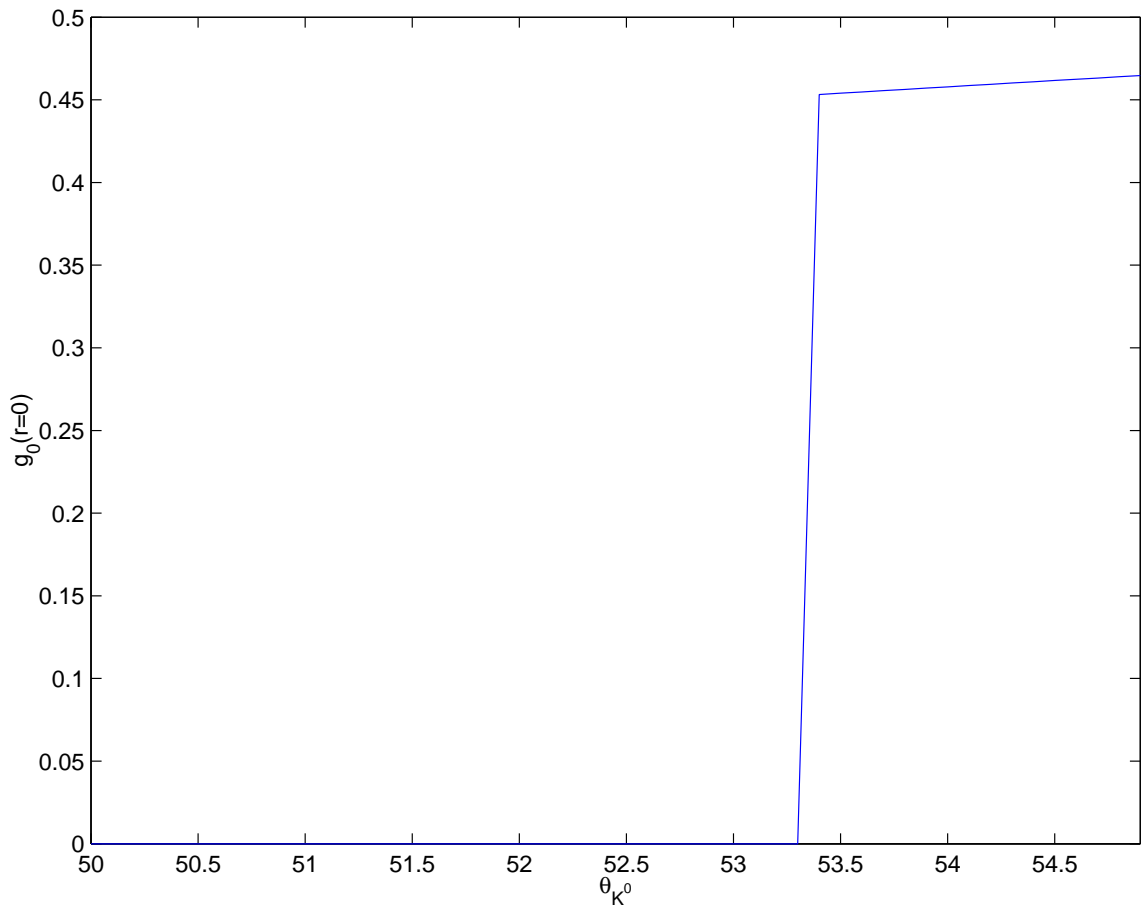}
\end{center}
\caption{ In this figure we plot the the value of the $\kplus$-condensate 
at the centre of the $\kzero$-string $g_0(r=0)$ ($g_0(r)$ and $\kplus$ are
related my Eq. (\ref{fourierexp})) as a function of the 
kaon condensation angle $\thetak$. From this graph, we can see that 
$\kplus$-condensation occurs at $\thetak > \thetacritone$.
Note that there is an abrupt point at which $\kplus$-condensation 
occurs inside the core. As described in the text, the transition to the 
$\kplus$-condensed core corresponds to a jump at $\thetacrit$. The finite
slope on the plot at this point is due to the discretization of the 
$\thetak$ variable. }
\label{gvsthetaphi4}
\end{figure}

\begin{figure}
\begin{center}
\vspace{1cm}
\epsfxsize=6.0in
\epsfbox[ 61 199 549 589]{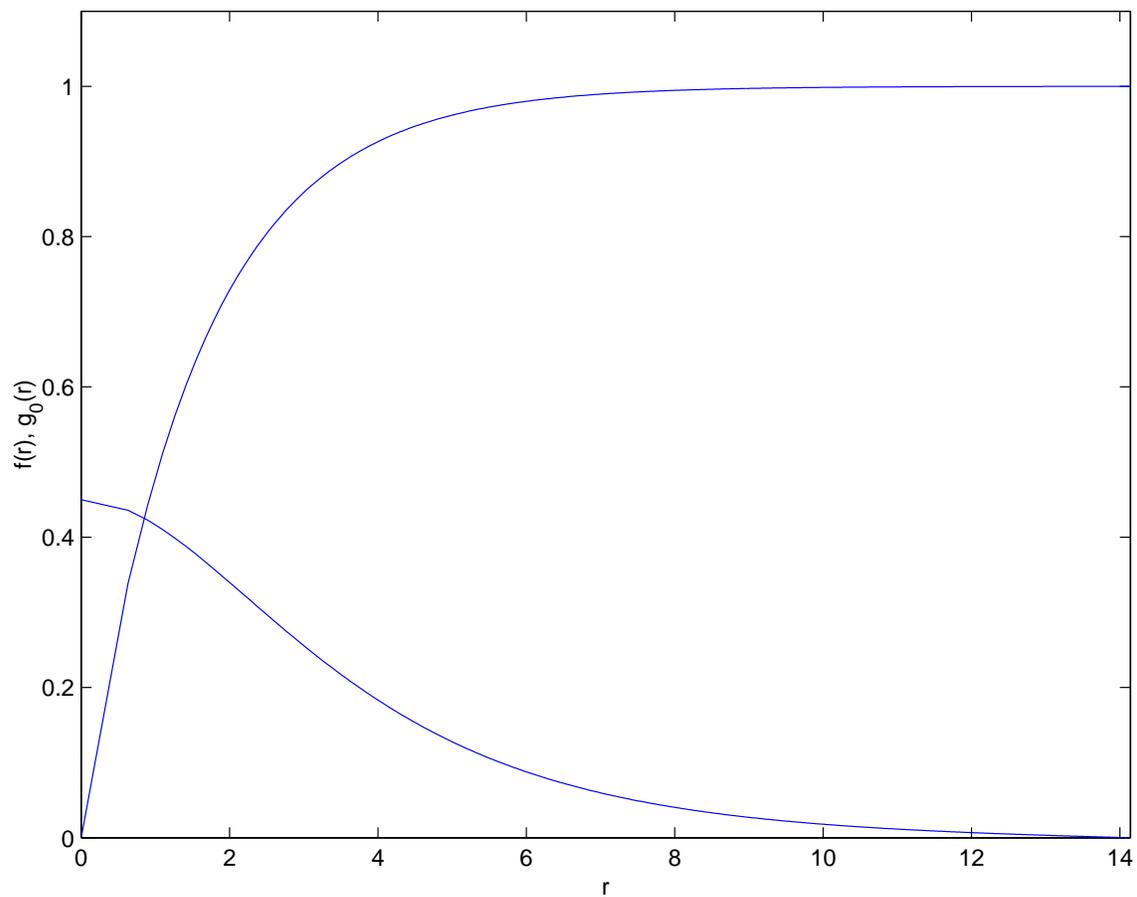}
\end{center}
\caption{ In this figure we plot the functions $f(r)$ (related to the 
$\kzero$-string by Eq. (\ref{k0string})) and $g_0(r)$ (related to the 
$\kplus$-condensate by Eq. (\ref{fourierexp})) as a function of the 
dimensionless rescaled coordinate $\rtil \equiv \vpi r / \mueff$.}
\label{k0k+vsr}
\end{figure}

\end{document}